\documentclass[pra,twoside,showpacs,superscriptaddress,twocolumn]{revtex4}
\usepackage{psfig}

\begin{document}

\title{Experimental asymmetric phase-covariant quantum cloning
       of polarization qubits}

\author{Jan Soubusta}
  \affiliation{Joint Laboratory of Optics of Palack\'{y} University and
     Institute of Physics of Academy of Sciences of the Czech Republic,
     17. listopadu 50A, 779\,07 Olomouc, Czech Republic}

\author{Lucie Bart{\r u}\v{s}kov\'{a}}
  \affiliation{Department of Optics, Faculty of Science, Palack\'y University,
     17.~listopadu 50, 772\,00 Olomouc, Czech~Republic}
     
\author{Anton\'{\i}n \v{C}ernoch}
  \affiliation{Joint Laboratory of Optics of Palack\'{y} University and
     Institute of Physics of Academy of Sciences of the Czech Republic,
     17. listopadu 50A, 779\,07 Olomouc, Czech Republic}
     
\author{Miloslav Du\v{s}ek}
  \affiliation{Department of Optics, Faculty of Science, Palack\'y University,
     17.~listopadu 50, 772\,00 Olomouc, Czech~Republic}

\author{Jarom{\'\i}r Fiur\'{a}\v{s}ek}
  \affiliation{Department of Optics, Faculty of Science, Palack\'y University,
     17.~listopadu 50, 772\,00 Olomouc, Czech~Republic}

\date{\today}

\begin{abstract}
We report on two optical realizations of the $1 \rightarrow 2$ asymmetric 
phase-covariant cloning machines for polarization states of single photons.
The experimental setups combine two-photon interference and tunable polarization 
filtering that enables us to control the asymmetry of the cloners. The first 
scheme involves a special unbalanced bulk beam splitter exhibiting different 
splitting ratios for vertical and horizontal polarizations, respectively. 
The second implemented scheme consists of a balanced fiber coupler where 
photon bunching occurs, followed by a free-space part with polarization 
filters. With this later approach we were able to demonstrate very high 
cloning fidelities which are above the universal cloning limit. 

\end{abstract}


\maketitle


\section{Introduction}

Optimal copying of quantum states is an important primitive in quantum information
processing \cite{Scarani05,Cerf06}. Since exact copying of unknown quantum states 
is forbidden due to the linearity of quantum mechanics \cite{Wootters82} this task
can be accomplished only approximately. A figure of merit commonly employed to quantify 
the performance of quantum cloners is the fidelity of the clones \cite{Scarani05,Cerf06}. 
Optimal cloning machines that maximize the cloning fidelity have been identified theoretically 
for a wide range of classes of input states and numbers of copies \cite{Scarani05,Cerf06}.
The universal quantum cloners copy all states from the underlying Hilbert space with 
the same fidelity \cite{Buzek96,Gisin97,Bruss98,Werner98}. 
Sometimes, however it is more beneficial to clone optimally only a certain subset of 
states. A particularly important example is the phase-covariant  quantum cloner 
\cite{Niu99,Bruss00,DAriano03,Fiurasek03} that optimally copies all qubits from the 
equator of the Bloch sphere, i.e. all balanced superpositions of the computational 
basis states. The advantage of such dedicated cloning machine is that it reaches
higher cloning fidelities than the universal machine.

Phase-covariant cloning represents an optimal individual eavesdropping
attack on BB84 quantum key distribution protocol \cite{Fuchs97,Cerf02}. 
In this context, the asymmetric cloning machines that
produce two copies with different fidelities \cite{Cerf02,Rezakhani05,Lamoureux05} 
are particularly important. Tuning the asymmetry 
of the cloning operation enables to control the trade-off between information 
on a secret cryptographic key gained by the eavesdropper and the amount of noise 
added to the copy  which is sent down the channel to the authorized receiver.

For potential applications in quantum communication, such as eavesdropping
on quantum key distribution, cloning of the quantum states of single photons 
is of great interest \cite{Scarani05,Cerf06}. Universal cloning of polarization states 
of single photons has been implemented experimentally using either stimulated 
parametric downconversion \cite{Linares02,DeMartini04} or bunching of photons 
on a balanced beam splitter \cite{Ricci04,Irvine04,Khan04}.  
Asymmetric universal cloning \cite{Zhao05} and symmetric $1\rightarrow 3$ 
phase-covariant cloning \cite{Sciarrino05} of photonic qubits have also been 
realized. Recently, we have experimentally demonstrated the optimal symmetric 
$1\rightarrow 2$ phase-covariant cloning of the polarization states of single 
photons \cite{Cernoch06,Soubusta07}. We have also implemented an all-fiber 
setup for optimal phase-covariant asymmetric cloning of qubits 
represented by single photons that can simultaneously propagate in two distinct 
single-mode optical fibers \cite{Bartuskova07}.

In the present paper we report on the experimental demonstration of the optimal 
asymmetric phase-covariant cloning of the polarization states of single photons. 
In contrast to our previous fiber-optics experiment our present approach does 
not rely on single-photon interference and we therefore do not have to stabilize 
any first-order interferometer. We have implemented two schemes both of which are 
extensions of setups utilized previously for symmetric phase-covariant 
cloning as described in Refs. \cite{Cernoch06,Soubusta07}.

The first setup involves an interference of two photons (signal, and an ancilla 
in a vertical polarization state) on a specially tailored unbalanced beam splitter, 
which ideally affects the optimal symmetric phase-covariant cloning operation 
\cite{Fiurasek03,Cernoch06}. The cloner is then made asymmetric by applying 
partial polarization filters to both clones. This filtering is realized by means 
of tilted glass plates, introducing different transmittances of the TE and TM 
polarization modes according to Fresnel equations. A second approach relies 
on the combination of optimal universal cloning and polarization filtration. 
The former is achieved by an interference of the two photons on a balanced beam 
splitter \cite{Ricci04,Irvine04,Khan04}.
We utilize advantageously a fiber coupler which allows us to reach very high 
visibility of the Hong-Ou-Mandel interference \cite{Hong87}. The two clones 
are then subjected to partial polarization filtration such that the output of 
the machine corresponds to that of the optimal phase-covariant cloner. 
With this latter scheme we observe cloning fidelities exceeding those achievable 
with any optimal universal cloning machine. We thus clearly demonstrate the 
advantage of phase-covariant asymmetric quantum cloners over the universal 
asymmetric cloner in cases of restricted sets of input states.

The paper is structured as follows. In Section \ref{sec_optimalPC} we briefly 
review the theory of the optimal asymmetric phase-covariant cloning operations. 
In Section \ref{sec_BS} we present the experimental implementation of the asymmetric 
cloner that relies on the two-photon interference on an unbalanced beam splitter 
followed by polarization filtration which tunes the asymmetry of the cloner. 
In Section \ref{sec_Hyb} we discuss an alternative scheme consisting of a sequence 
of the optimal universal cloner based on photon bunching in a balanced fiber 
coupler followed again by appropriate polarization filtration. Finally, 
section \ref{sec_conc} contains a brief summary of the main results and a comparison of
the two schemes.


\section{Optimal phase-covariant cloning}\label{sec_optimalPC}

We are interested in copying of a polarization  state of  a single photon $|\psi \rangle$. 
This single-qubit state can be conveniently parametrized by two 
Euler angles $\theta$ and $\phi$,
\begin{equation}
   |\psi \rangle = \cos{\frac{\theta}{2}} |V \rangle + e^{i \phi} \sin{\frac{\theta}{2}} |H \rangle.
\end{equation}
Here the two orthogonal computational basis states $|V\rangle$ and $|H\rangle$ 
represent the vertical and horizontal linear polarization states, respectively. 
In this paper we focus on the cloning of the polarization states situated on 
the equator of the Bloch sphere $(\theta= {\pi \over 2})$,
\begin{equation}\label{EQ_psi} 
   |\psi \rangle = {1 \over \sqrt{2}} \big(|V \rangle + e^{i \phi} |H \rangle\big).
\end{equation}
The optimal asymmetric phase-covariant cloning transformation reads \cite{Bartuskova07},
\begin{equation}\label{EQ_transf}
 \begin{array}{ll}
   |V \rangle|V \rangle \rightarrow |V\rangle|V \rangle ,\\
   |H \rangle|V \rangle \rightarrow \sqrt{q} \, |V\rangle|H \rangle 
                                 + \sqrt{1-q}\, |H\rangle|V \rangle ,
 \end{array}
\end{equation}
where $q \in [0,\,1]$ is the asymmetry parameter. Note that this unitary transformation 
requires only two qubits, the signal whose state we want to clone and an ancilla qubit 
(a blank copy) prepared in a fixed state $|V\rangle$.
The second line of Eq. (\ref{EQ_transf}) means creation of a superposition of the input state 
with a state where the two photons have been exchanged. Such states are naturally produced 
by a beam splitter with splitting ratio depending on the asymmetry parameter $q$.   

The quality of the clones is quantified by their fidelity,
which is defined as the overlap of each clone state with the original state (\ref{EQ_psi}). 
The fidelities of the two clones produced by the optimal asymmetric phase-covariant cloning
transformation (\ref{EQ_transf}) read,
\begin{equation}\label{EQ_F1F2_pc}
   F_1 = \frac{1}{2} \left( 1+ \sqrt{1-q} \right), \qquad 
   F_2 = \frac{1}{2} \left( 1+ \sqrt{q} \right).
\end{equation}
In the case of symmetric cloning ($q=1/2$) both fidelities have the same value 
$F_{\rm sym,pc} \approx 85.4\%$. For comparison we give also the clone fidelities 
achievable by the optimal universal asymmetric cloning \cite{Cerf06},
\begin{equation}\label{EQ_F1F2_univ}
   F_{\rm u1} = 1 -{(1-p)^2 \over 2(1-p+p^2)}, \quad 
   F_{\rm u2} = 1 -{   p^2  \over 2(1-p+p^2)},
\end{equation}
where parameter $p \in [0,\,1]$ controls the asymmetry of the two clones.
An universal cloner copies all states (not only the equatorial ones) with 
the same fidelities $F_{\rm u1}$ and $F_{\rm u2}$.
The fidelity of the symmetric universal cloner ($p=1/2$) reads $F_{\rm sym, univ} 
\approx 83.3\%$ \cite{Buzek96,Gisin97}.


\section{Free space realization with a special beam splitter}\label{sec_BS}

\begin{figure}[h]
\centerline{\psfig{figure=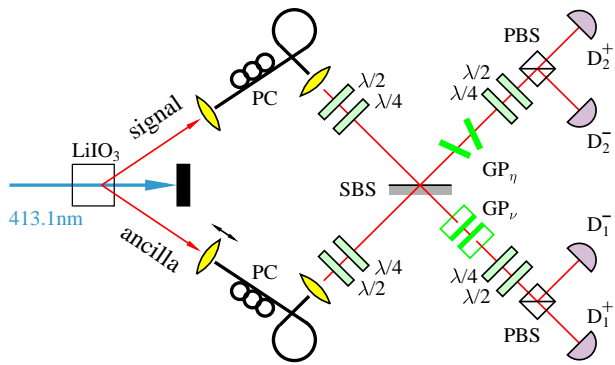,width=\linewidth}}
\caption{(Color online) Scheme of the cloning setup based on the special beam splitter 
  and polarization dependent
  losses. PC - polarization controller, SBS - special beam splitter, 
  GP$_{\eta}$, GP$_{\nu}$ - polarization dependent losses, 
  PBS - polarizing cube beam splitter, 
  $\lambda/2, \; \lambda/4$ - wave plates, 
  D - detector.
  \label{setup82asym}}
\end{figure}

The first setup for the optimal asymmetric phase-covariant cloning of polarization 
states of single photons is shown in Fig.~\ref{setup82asym}. This setup is based 
on an interference of two photons on a special unbalanced beam splitter (SBS) 
with splitting ratios different for vertical and horizontal polarizations. The 
interference on SBS is followed by polarization filtration performed on each 
output port of the beam splitter. The polarization filters are realized by tilted 
glass plates (GP), where the tilt angle determines the ratio of transmittances for 
the horizontal and vertical polarizations. As we shall show below, by tilting the 
plates we are able to control the asymmetry of the cloner. The device operates 
in the coincidence basis and successful cloning is heralded by the presence of
a single photon in each output port of the cloning machine. In practice, we 
postselect only the cases when we observe coincidence between photon
detections in the upper and lower output arms. All other events are discarded.

Let us describe the experimental setup in more details. 
A non-linear crystal of LiIO$_3$ is pumped by cw Kr$^+$ laser at 413 nm to produce 
pairs of photons in the type I process of spontaneous parametric down conversion. 
Photons comprising each pair exhibit tight time correlations and are horizontally 
polarized. The photons are coupled into single mode fibers that serve as spatial 
filters. The polarization controllers (PC) are adjusted such as to ensure horizontal 
linear polarization of the two photons at the outputs of the fibers. The polarization 
state of each photon is set by means of half- and quarter-wave plates 
$(\lambda/2, \lambda/4)$. 
The photon in the upper  arm represents the signal qubit whose state should be cloned. 
The other photon serves as the ancilla and its polarization state is fixed to 
$|V\rangle$, c.f. Eq. (\ref{EQ_transf}). Both photons enter the special beam splitter 
which forms the Hong-Ou-Mandel interferometer \cite{Hong87}. 
The pairs of tilted glass plates introduce different amplitude transmittances 
for horizontal and vertical polarizations, ($\eta_V$, $\eta_H$ for GP$_{\eta}$; 
and $\nu_V, \nu_H$ for GP$_{\nu}$). It is convenient to define the 
intensity transmittance ratios, 
\begin{equation}
\Sigma_\eta=\left( \frac{\eta_V}{\eta_H}\right)^2, \qquad 
\Sigma_\nu=\left( \frac{\nu_V}{\nu_H}\right)^2. 
\end{equation}
GP$_{\eta}$ dominantly attenuates vertical polarization, hence $\Sigma_\eta\le 1$, 
while GP$_{\nu}$ imposes higher losses for horizontal polarization, and $\Sigma_\nu\ge 1$. 
We use two glass plates in each arm to reach higher transmittance ratios for the 
two polarizations. Moreover, since the two plates are tilted in opposite directions,
the beams are not transversally displaced by the filtration.

The transformation introduced by the setup shown in Fig.~\ref{setup82asym} can be 
written in the form
\begin{equation}
\begin{array}{rl}
|V \rangle_{\rm sig} |V \rangle_{\rm anc} \rightarrow 
   & {\eta_V \nu_V} (r_V^2 - t_V^2 ) |V\rangle_1|V \rangle_2, \\
|H \rangle_{\rm sig} |V \rangle_{\rm anc}  \rightarrow 
   & {\eta_H \nu_V} r_V r_H |V\rangle_1|H \rangle_2 \\
   & - {\eta_V \nu_H} t_V t_H |H\rangle_1 |V \rangle_2, 
   \label{HVfirst}
\end{array}
\end{equation}
where $r_V, r_H; \, t_V, t_H$ are the (fixed) real amplitude reflectances and 
transmittances of the SBS. We use notation  $R_j = r_j^2$ and $T_j = t_j^2$, $j=H,V$, 
for intensity reflectances and transmittances  and we have $R_j + T_j = 1$  
for a lossless beam splitter.

\begin{figure}
\centerline{\psfig{figure=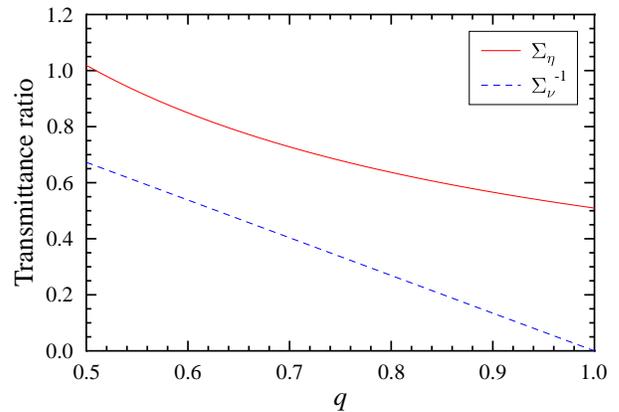,width=0.95\linewidth}}
\caption{(Color online) Transmittance ratios $\Sigma_\eta$ and $\Sigma_\nu^{-1}$ 
   for the asymmetric cloner with the special unbalanced beam splitter. 
   Plotted dependences were calculated according to Eq. (\protect{\ref{EQ_eta_82}}) 
   using experimentally determined parameters of SBS: $R_V=75.8\%$ and $R_H=17.9\%$.
   \label{Teorie_82}}
\end{figure}

Mapping (\ref{HVfirst}) becomes equivalent to the unitary cloning transformation 
(\ref{EQ_transf}) up to an overall prefactor representing the amplitude of the 
probability of success of the cloning if the following two conditions are 
satisfied simultaneously,
\begin{eqnarray}
  \eta_H \nu_V r_V r_H  &=& \sqrt{q} \, \eta_V \nu_V (r_V^2 - t_V^2), \nonumber \\
 - \eta_V \nu_H t_V t_H  &=& \sqrt{1-q} \, \eta_V \nu_V (r_V^2 - t_V^2).
\end{eqnarray}
After some algebra we obtain the transmittance ratios of the polarization filters 
expressed as functions of the asymmetry parameter $q$,
\begin{eqnarray}\label{EQ_eta_82}
 \Sigma_{\eta} &=& {R_V R_H \over (2R_V-1)^2}{1 \over q}, \nonumber \\
 \Sigma_{\nu}  &=& {(1-R_V)(1-R_H) \over (2R_V-1)^2} \frac{1}{1-q}.
\end{eqnarray}
We have chosen the splitting ratios of the SBS such that symmetric cloning could be realized
without any further polarization filtration. If we set $\Sigma_\eta=\Sigma_\nu=1$ and $q=1/2$
in Eq. (\ref{EQ_eta_82}) we obtain $R_V=\frac{1}{2}(1+\frac{1}{\sqrt{3}}) \approx 78.9\%$ 
and $R_H=1-R_V$ \cite{Cernoch06}.  The experimentally determined parameters 
of the custom-made SBS manufactured by Ekspla read $R_V=75.8\%$ and $R_H=17.9\%$ 
which is close to the desired values. 
Figure~\ref{Teorie_82} shows the theoretical dependence of the transmittance ratios on $q$
calculated according to Eq. (\ref{EQ_eta_82}) using the experimentally determined values of 
$R_H$ and $R_V$. In particular, note that the symmetric operation would be achieved for 
$\Sigma_\nu^{-1} = 0.67$ and $\Sigma_\eta = 1.02$.
The probability of success of the cloning is given by 
$P_{\rm succ}^{\rm SBS} = \eta_V^2 \nu_V^2 (r_V^2 - t_V^2)^2$. It can be shown that 
$P_{\mathrm{succ}}$ is highest for the symmetric cloner and decreases with increasing 
asymmetry, because the losses introduced by polarization filters are increasing. Comparison of 
the measured and theoretically attainable $P_{\mathrm{succ}}$ is given at the end of 
Sec.~\ref{sec_Hyb} for both setups.

As we already mentioned, the cloning procedure is successful only if there is one 
photon in each output arm of the device. The performance of the cloning machine is 
probed by polarization analysis of the two clones. The setting of wave plates at 
the output is inverse with respect to the signal photon preparation. This means that 
the photons with the same polarization as the signal photon are transmitted through 
the PBS to the detector D$^+$ whereas the photons with orthogonal polarization are 
reflected to the detector D$^-$. This allows us to infer the cloning fidelities from 
the four measured coincidence rates $C^{\pm \pm}$ between detectors at the two output 
arms. For instance, coincidence rate $C^{++}$ represents the number of simultaneous 
clicks of detectors D$_1^+$ and D$_2^+$ per second and the other coincidence rates 
are determined similarly. The fidelities of the clones are calculated as the ratio 
of coincidences corresponding to the projection of the first (second) clone onto the 
input state and the sum of all coincidences 
$C_{\mathrm{sum}} = {C^{++} + C^{+-} + C^{-+} + C^{--}}$,
\begin{eqnarray}
F_1 = \frac{C^{++} + C^{+-}}{C_{\mathrm{sum}}} , \qquad
F_2 = \frac{C^{++} + C^{-+}}{C_{\mathrm{sum}}} .
\end{eqnarray}
The probability of success of the device is determined as a fraction of the sum of 
all measured coincidences to the total number of the photon pairs entering the 
cloner $C_{\mathrm{tot}}$, $P_{\rm succ} = C_{\mathrm{sum}}/C_{\mathrm{tot}}$.

\begin{figure}
\centerline{\psfig{figure=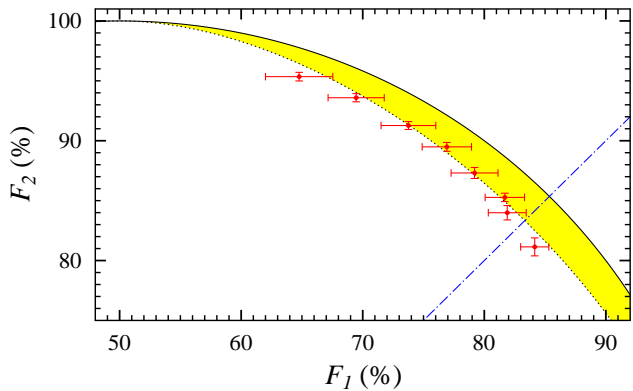,width=\linewidth}}
\caption{(Color online) 
  Fidelities $F_2$ vs $F_1$ of clones measured with the setup based on the special 
  beam splitter and polarization dependent losses. The full line denotes the theoretical 
  limit for the fidelities of the phase-covariant cloner, the dotted line shows the limit 
  of the universal cloner. The dashed line shows the symmetric
  line ($F_1=F_2$).
  \label{F1F2_82}}
\end{figure}

\begin{table}
\centerline{
\begin{tabular}{cccccc} \hline
 $q$ & 
   $\Sigma_{\nu}^{-1}$ & 
   $\Sigma_{\eta}$ &
     $F_1$ [\%] & 
     $F_2$ [\%] & 
       $P_{\rm succ}^{\mathrm{SBS}}$ [\%] \\ \hline \hline
0.93 & 0.10 & 0.55 &   $64.8\pm 2.8$ &   $95.4 \pm 0.4$ &    $2.9 \pm 0.1$ \\ 
0.85 & 0.20 & 0.60 &   $69.5\pm 2.3$ &   $93.6 \pm 0.3$ &   $10.3 \pm 0.1$ \\ 
0.78 & 0.30 & 0.66 &   $73.8\pm 2.3$ &   $91.3 \pm 0.3$ &   $13.8 \pm 0.1$ \\ 
0.70 & 0.40 & 0.73 &   $76.9\pm 2.0$ &   $89.5 \pm 0.4$ &   $17.4 \pm 0.2$ \\ 
0.63 & 0.50 & 0.81 &   $79.2\pm 1.9$ &   $87.3 \pm 0.5$ &   $18.8 \pm 0.2$ \\ 
0.55 & 0.60 & 0.92 &   $81.7\pm 1.6$ &   $85.3 \pm 0.4$ &   $20.0 \pm 0.2$ \\ 
0.51 & 0.66 & 1.00 &   $81.9\pm 1.6$ &   $84.0 \pm 0.6$ &   $23.6 \pm 0.2$ \\ 
-    & -    & -    &   $84.2\pm 1.2$ &   $81.1 \pm 0.8$ &   $28.8 \pm 0.2$ \\ \hline
\end{tabular}}
\caption{Table of the setting of glass plates, measured fidelities and 
   $P_{\rm succ}^{\mathrm{SBS}}$ 
   for the specified asymmetry parameter $q$ for the setup based on the SBS. 
   The last row represents the measurement without glass plates. \label{table82}}
\end{table}

We selected seven representative values of asymmetry parameter $q$ and set the angles 
of the GP$_\eta$ and GP$_\nu$ accordingly. Then we measured clone fidelities for 
a set of nine states $\frac{1}{\sqrt{2}}(|V\rangle+e^{ik\pi/4}|H\rangle)$,
 $k=-4,\ldots, 4$, located on the equator of the Bloch 
sphere. These states span over circular and diagonal linear polarization states. 
Resulting mean fidelities averaged over the nine states are plotted in Fig.~\ref{F1F2_82} 
and listed in Tab.~\ref{table82}. 
The statistical errors were calculated from 10 ten-second measurements. They reach values 
$\sim 1\%$ in the symmetric case. For higher degrees of asymmetry, the polarization 
filtration resulted in higher losses, leading to decrease of the success probability 
and increase of statistical errors of fidelities up to $\sim 3\%$. 
The last row of Tab.~\ref{table82} represents the measurement without glass plates. 
It shows the intrinsic asymmetric operation of the SBS. The use of the polarization 
filters enables to approach the symmetric operation and to go further to tune the 
asymmetry up to markedly asymmetric cloning.

The statistical error of mean fidelity $F_1$ is greater than the error of $F_2$. 
This is due to more pronounced oscillations of $F_1$ when scanning over the 
equatorial states. This effect is caused by residual uncompensated phase shifts 
induced by the special beam splitter. The reflected and transmitted photons 
acquire different phase shifts. Figure~\ref{F1F2_82} shows the comparison of our 
measurements with the theoretical limits of the universal asymmetric cloner 
(\ref{EQ_F1F2_univ}) and of the phase-covariant asymmetric cloner (\ref{EQ_F1F2_pc}). 
All measured points are very close to the universal asymmetric cloning limit 
but do not reach the theoretical phase-covariant cloning limit on fidelity.
The main effect that reduces the fidelity of the two clones and prevents us to 
surpass the universal cloning limit is the non-ideal overlap of the spatial modes 
of the two photons on the SBS.


\section{Hybrid free-space and fiber setup}\label{sec_Hyb}

\begin{figure}[h]
\centerline{\psfig{figure=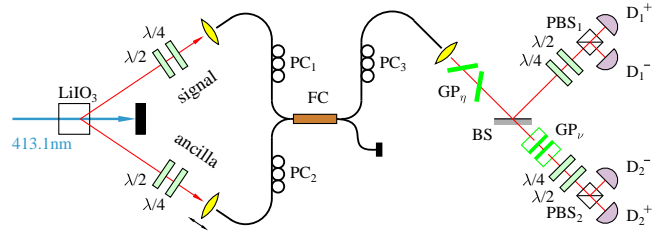,width=\linewidth}}
\caption{(Color online) Scheme of the hybrid cloning setup. 
  FC - fiber coupler, BS - nonpolarizing beam splitter, 
  PBS - polarizing beam splitter, PC - polarization controller, 
  $\lambda/2, \; \lambda/4$ - wave plates, 
  GP$_{\eta}$, GP$_{\nu}$ - polarization dependent losses,
  D - detector.
  \label{setupHyb}}
\end{figure}

In order to increase the cloning fidelity we have built an alternative setup which 
combines advantages of both fiber and free space propagation, see Fig.~\ref{setupHyb}. 
Fiber coupler (FC) ensures perfect overlap of spatial modes of signal and ancilla 
photons. The free space part allows to use simple encoding of information 
into the polarization states of the photons. We can use wave plates $(\lambda/2, 
\lambda/4)$ and polarizing beam splitters to prepare arbitrary input polarization states
and to perform projective measurements in arbitrary basis. 
This experimental scheme is based on the universal cloner, i.e. interference 
of two photons on a balanced beam splitter \cite{Ricci04,Irvine04,Khan04}, 
which is modified by the state filtering. The polarization filters GP$_{\eta}$ 
and GP$_{\nu}$ ensure implementation of phase-covariant cloning transformation, 
compensate for non-ideal splitting ratio of the BS and allow to tune the asymmetry 
of the cloner. The device again operates in the coincidence basis
and the cloning is successfully accomplished if a single photon is detected 
in each output port of the cloner.
The transformation realized by the whole device can be written as,
\begin{equation}
\begin{array}{rl}\label{EQ_operace_Hyb}
|V \rangle_{\rm sig} |V \rangle_{\rm anc} \rightarrow &  
  2rt \eta_V^2 t_V r_V \nu_V |V\rangle_1|V \rangle_2, \\
|H \rangle_{\rm sig} |V \rangle_{\rm anc} \rightarrow & 
  rt \eta_V\eta_H (t_V r_H \nu_V |H\rangle_1|V \rangle_2\\
                 & + t_H r_V \nu_H |V\rangle_1|H \rangle_2),
\end{array}
\end{equation}
where coefficients $r$ and $t$ represent reflectance and transmittance of the FC;
$r_{j}$ and $t_{j}$, $j=V,H$, denote reflectance and transmittance of the BS,
which are slightly polarization dependent 
$(R_V = 50.9\%$ and $R_H = 46.6\%)$.

\begin{figure}
\centerline{\psfig{figure=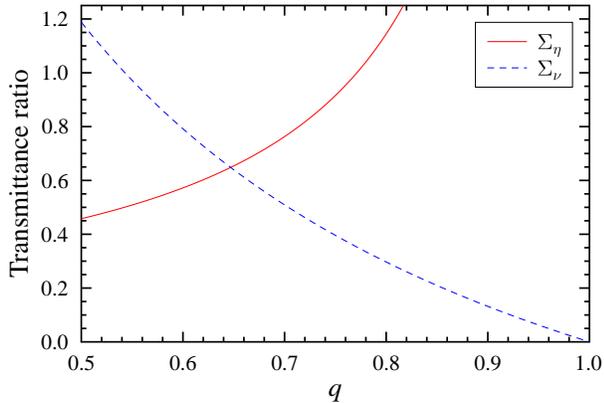,width=0.95\linewidth}}
\caption{(Color online) Transmittance ratios $\Sigma_\eta$ and $\Sigma_\nu$ for 
   hybrid asymmetric cloning setup.
   Plotted dependences were calculated according to Eq. (\protect{\ref{EQ_eta_Hyb}}) 
   using experimentally determined parameters of BS: $R_V = 50.9\%$ and $R_H = 46.6\%$. 
   \label{Teorie_Hyb}}
\end{figure}

Similarly as for the previous scheme, Eq. (\ref{EQ_operace_Hyb}) becomes equivalent 
to the target unitary cloning transformation (\ref{EQ_transf}) if the following 
relations hold,
\begin{eqnarray}
\eta_V \eta_H t_V r_H \nu_V &=& \sqrt{1-q} \, 2 \eta_V^2 t_V r_V \nu_V, \nonumber \\
\eta_V \eta_H t_H r_V \nu_H &=& \sqrt{q} \, 2 \eta_V^2 t_V r_V \nu_V.
\end{eqnarray}
After some algebra we arrive at the dependence of the transmittance ratios of the 
polarization filters on the setup parameters and the asymmetry parameter $q$,
\begin{equation}\label{EQ_eta_Hyb}
 \Sigma_\eta  = {R_H \over R_V}{1 \over 4(1-q)}, \quad 
 \Sigma_\nu  = {R_V(1-R_H) \over R_H(1-R_V)}{1-q \over q}.
\end{equation}
Dependences of $\Sigma_\eta$ and $\Sigma_\nu$ on $q$ calculated for the parameters 
of our setup are plotted in Fig~\ref{Teorie_Hyb}.

The measurement routine starts with an adjustment of the HOM interference dip in 
the fiber coupler FC. In this preliminary stage two outputs of the FC are connected 
directly to the detectors and the overlap of the two photons is maximized finding 
a minimum of the coincidence counts. Optimal overlap of the polarization states on 
the FC is achieved by adjusting polarization controllers PC$_1$ and PC$_2$.
Then one output of the FC is directed to the free space part of the setup. 
The last polarization controller PC$_3$ is used to compensate polarization 
transformation induced in the fibers. The BS splits 
the photon pair into two paths with probability $1 \over 2$. GP$_{\eta}$ and 
GP$_{\nu}$ are tilted to provide demanded polarization state filtration.  
Input polarization states and measurement bases are set by half- and quarter-wave plates.

\begin{figure}
\centerline{\psfig{figure=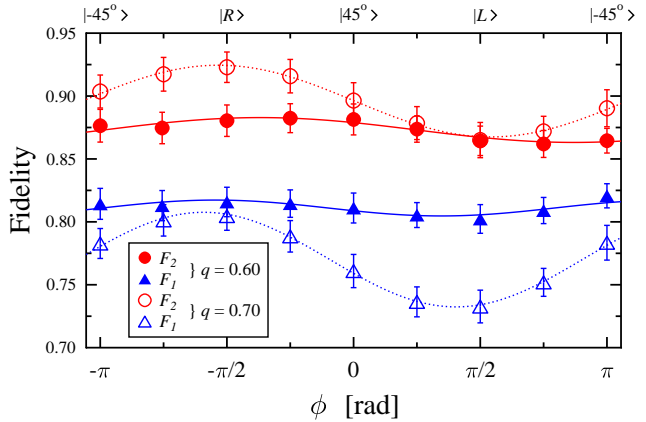,width=\linewidth}}
\caption{(Color online) Fidelities $F_1$ and $F_2$ of cloning of equatorial qubits with the hybrid 
        cloning setup for two asymmetry parameters, $q=0.60$ and $q=0.70$. The top 
        axis shows the signal qubit state corresponding to the phase $\phi$
        (see Eq.~(\protect{\ref{EQ_psi}})). 
   \label{equatorHyb}}
\end{figure}

As in the previous section we performed cloning of nine polarization 
states of a signal qubit distributed over the equator of the Bloch sphere. 
Figure~\ref{equatorHyb} shows two typical examples of the experimentally 
measured fidelities for the equatorial qubits $(q=0.60$ not oscillating, 
$q=0.70$ the most oscillating one). Statistical error bars were determined 
from 10 twenty~second measurement periods. Due to the fact, that the oscillations 
have the sinusoidal character and for both fidelities the sinusoids have the same 
phase, we suppose that we did not set the ancilla photon polarization exactly on 
the pole of the Bloch sphere. Higher oscillations lead to greater errors of the 
mean fidelities.

Note that any cloner can be converted by a twirling operation \cite{Cerf06} to 
a truly phase-covariant cloner whose cloning fidelity does not depend on the 
input state and is equal to the mean fidelity of the original cloner. 
The twirling consists in application of the random phase shift operation 
$U(\vartheta)=|V\rangle\langle V|+e^{i\vartheta}|H\rangle\langle H|$ to the input
state and the inverse operation $U(-\vartheta)$ to each of the clones. The phase shift
$\vartheta$ is selected randomly from the interval $[0,2\pi]$. In the present
implementation, the twirling could be performed by using additional wave-plates. 

\begin{figure}
\centerline{\psfig{figure=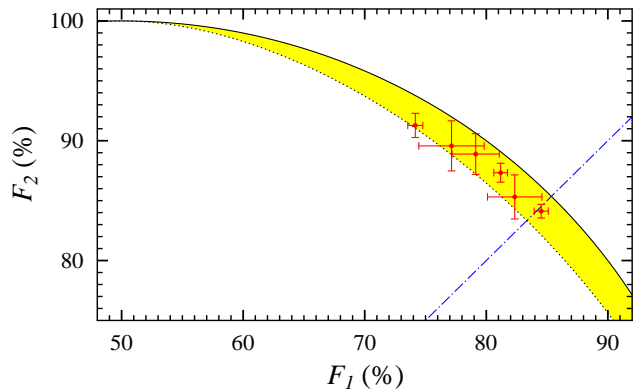,width=\linewidth}}
\caption{(Color online) Fidelities $F_2$ vs $F_1$ of clones measured with the hybrid setup. 
  The full line denotes the theoretical limit for the fidelities of the 
  phase-covariant cloner, the dotted line shows the limit of the universal 
  cloner. The dashed line shows the symmetric line ($F_1=F_2$).
  \label{F1F2_Hyb}}
\end{figure}

\begin{table}
\centerline{
\begin{tabular}{cccccc} \hline
 $q$ & 
   $\Sigma_{\eta}$ &
   $\Sigma_{\nu}$ & 
     $F_1$ [\%] & 
     $F_2$ [\%] & 
       $P_{\rm succ}^{\mathrm{Hyb}}$ [\%] \\ \hline \hline
0.75 & 1.00 & 0.39 &   74.2 $\pm$ 0.6 &   91.3 $\pm$ 1.0 &   6.2 $\pm$ 0.2 \\ 
0.70 & 0.83 & 0.50 &   77.1 $\pm$ 2.7 &   89.6 $\pm$ 2.1 &   4.8 $\pm$ 0.2 \\ 
0.65 & 0.71 & 0.63 &   79.1 $\pm$ 1.9 &   88.9 $\pm$ 1.7 &   4.8 $\pm$ 0.1 \\ 
0.60 & 0.63 & 0.78 &   81.2 $\pm$ 0.6 &   87.3 $\pm$ 0.8 &   5.0 $\pm$ 0.1 \\ 
0.55 & 0.56 & 0.96 &   82.3 $\pm$ 2.2 &   85.3 $\pm$ 1.8 &   4.4 $\pm$ 0.2 \\ 
0.50 & 0.50 & 1.17 &   84.5 $\pm$ 0.6 &   84.1 $\pm$ 0.6 &   4.2 $\pm$ 0.1 \\ \hline
\end{tabular}}
\caption{Table of the setting of glass plates, measured fidelities and 
   $P_{\rm succ}^{\mathrm{Hyb}}$ for the specified asymmetry parameter 
   $q$ of the hybrid setup. \label{tableHyb}}
\end{table}

The relevant parameters of the cloner are thus the mean cloning fidelities which fully 
quantify its performance.  The mean fidelities are shown in Fig.~\ref{F1F2_Hyb} and 
are also listed in Table~\ref{tableHyb}. As can be seen the resulting mean fidelities 
are above the universal cloning limit for all asymmetries. 
Note that due to technical limitations on achievable $\Sigma_\eta$ and $\Sigma_\nu$ 
we can reach only moderate asymmetries $q \in [0.50,0.75]$.

\begin{figure}
\centerline{\psfig{figure=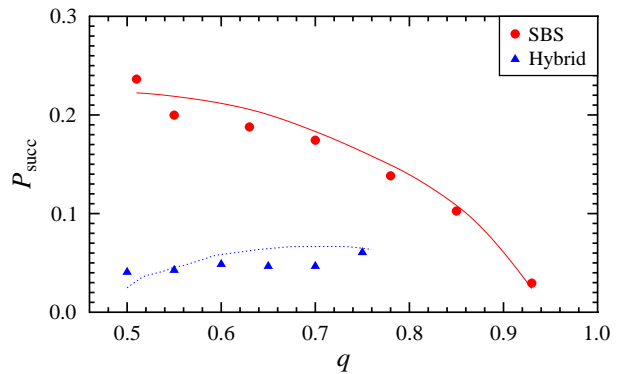,width=0.95\linewidth}}
  \caption{(Color online) Probability of success: the setup based on the special beam splitter 
  (circles), hybrid setup (triangles). The errors are smaller than the symbols
  shown. The lines represent theoretical dependences calculated from measured
  transmittances of the tilted glass plates. 
  \label{Fig_Psucc}}
\end{figure}

From the relations (\ref{EQ_operace_Hyb}) we can also determine the probability of 
success of the hybrid setup, $P_{\rm succ}^{\rm Hyb} = (2rt \eta_V^2 t_V r_V \nu_V)^2$.
For ideal symmetric cloner we obtain  $P_{\rm succ}^{\rm Hyb}={1 \over 16}$ which 
should be compared with $P_{\mathrm{succ}}^{\mathrm{SBS}}=\frac{1}{3}$ achieved by the
setup discussed in Sec.~\ref{sec_BS}. The hybrid setup exhibits lower probability of success 
mainly because there are two post-selection steps. First, the signal and ancilla 
photon must leave the FC together by the selected output fiber (the upper one in 
Fig.~\ref{setupHyb}). Second, there must be one photon in each output arm of the 
bulk BS. For completeness, we plot the measured probabilities of success of both 
cloning setups in Fig.~\ref{Fig_Psucc}.

\section{Conclusions}\label{sec_conc}

In this paper we described two experimental setups proposed to realize optimal 
asymmetric phase-covariant cloning of single-photon polarization qubits.
We characterized the real experimental operation of both setups and compared  
their performances and limitations.  
The cloning is based on interference of the signal photon with the ancilla photon 
on a beam splitter or fiber coupler followed by polarization filtration on the 
outputs. The implemented cloning machines operate in the coincidence basis and 
a successful operation of the device is heralded by detection of a single photon 
in each output arm.
An important feature of both experimental setups is that the polarization 
filtering allows to tune the asymmetry of the cloning 
operation. Moreover, the same polarization filtering is used to compensate 
imperfections of beam splitters whose splitting ratios slightly differed 
from the desired ones.

The first setup relies on a special unbalanced beam splitter with different
transmittances for vertical and horizontal polarizations. The main advantage 
of this setup is that we can tune the asymmetry of cloning in a broad range. 
However, the imperfect overlap of the spatial modes of the photons on the bulk
beam splitter limits the achievable fidelity of the clones and prevents us 
from surpassing the limit of optimal universal asymmetric cloning with this 
approach.
The second setup is based on the fiber coupler ensuring practically perfect 
overlap of spatial modes. 
With this second approach we were able to achieve very high mean cloning
fidelities exceeding the maximum fidelities obtainable by universal cloners.
To the best of our knowledge, this is the first experiment where universal cloning limit
has been surpassed for asymmetric cloning of equatorial polarization states of single
photons.
The price to pay for the fidelity improvement is a smaller probability of 
success of this latter scheme and also somewhat narrower accessible asymmetry 
range. 


\acknowledgments

This research was supported by the projects LC06007, 1M06002 and MSM6198959213 
of the Ministry of Education of the Czech Republic and by the EU project SECOQC 
(IST-2002-506813).


\end{document}